\pdfoutput=1

\documentclass{iaus}
\usepackage{graphicx}

\title[Obscuring and feeding BH with nuclear star clusters] %% give here short title %%
{Obscuring and feeding supermassive black holes with evolving nuclear star clusters}

\author[M.~Schartmann et al.]   %% give here short author list %%
{M.~Schartmann$^{1,2}$
 \and A.~Burkert$^{1,2,3}$
 \and M.~Krause$^{1,2}$
 \and M.~Camenzind$^4$
 \and K.~Meisenheimer$^5$
 \and R.~I.~Davies$^1$
}

\affiliation{$^1$Max-Planck-Institut f\"ur extraterrestrische Physik, Giessenbachstrasse, 85748 Garching, Germany\\ email: {\tt schartmann@mpe.mpg.de} \\[\affilskip]
$^2$Universit\"atssternwarte M\"unchen, Scheinerstrasse 1, 81679 M\"unchen, Germany\\[\affilskip]
$^3$Max-Planck-Fellow\\[\affilskip]
$^4$ZAH - Landessternwarte Heidelberg, K\"onigstuhl 12, 69117 Heidelberg, Germany\\[\affilskip]
$^5$Max-Planck-Institut f\"ur Astronomie, K\"onigstuhl 17, 69117 Heidelberg, Germany
}

\pubyear{2009}
\volume{267}  %% insert here IAU Symposium No.
\pagerange{119--126}
\jname{Co-Evolution of Central Black Holes and Galaxies}
\editors{B.M.\ Peterson, R.S.\ Somerville, \& T.\ Storchi-Bergmann, eds.}
\begin{document}

\maketitle

\begin{abstract}
Recently, high resolution observations with the help of the near-infrared
adaptive optics integral field spectrograph SINFONI at the VLT proved the
existence of massive and young nuclear star clusters in the centres of a
sample of Seyfert galaxies. 
With the help of high resolution hydrodynamical
simulations with the {\sc PLUTO}-code, we follow the evolution of
such clusters, especially focusing on mass and energy feedback from young stars. 
This leads to a filamentary inflow of gas on
large scales (tens of parsec), whereas a turbulent and very dense disc builds up on
the parsec scale. Here, we concentrate on the long-term evolution of the nuclear disc in NGC\,1068
with the help of an effective viscous disc model, using the mass input from the large
scale simulations and accounting for star formation in the disc. 
This two-stage modelling enables us to connect 
the tens of parsec scale region (observable with SINFONI) with the parsec scale environment 
(MIDI observations). At the current age of the
nuclear star cluster, our simulations predict disc sizes of the order of 0.8 to 
0.9\,pc, gas masses of $10^6\,M_{\odot}$ and mass transfer rates through the
inner boundary of $0.025\,M_{\odot}$/yr in good agreement with values derived 
from observations.
\end{abstract}

\firstsection % if your document starts with a section,
\section{Introduction}

Nuclear activity is an important phase in the evolution of galaxies. Whenever 
enough gas is fed onto the central supermassive black holes, galaxies become 
{\it active} and their nuclear region lights up significantly, compared to 
non-active phases of their evolution. 
The infalling material builds up a fast rotating accretion disc, which 
heats up viscously. The emerging UV/optical radiation illuminates a gas
and dust reservoir -- the so-called {\it molecular torus}. Given the high
opacity of the dust in this wavelength range, the torus morphology is 
able to geometrically unify two observed classes of active galaxies: type\,1 
objects, where the torus is viewed face-on (and all of the characteristics of the
central region are visible) and type\,2 objects (torus viewed edge-on), where most of 
the light from the accretion disc is absorbed and reemitted in the infrared regime.
This is the essence of the {\it Unified Scheme of Active Galactic Nuclei}. 
Despite its general success when probed with observational data, the nature of the
obscuring gas and dust reservoir -- the torus, as well as  
the triggering mechanism of such phases are still 
a matter of active research. To assess the final stages of mass transport
through the central tens of parsec region, nearby objects are needed,  
due to the limitations on resolution posed by currently available 
instruments. Therefore, Seyfert galaxies are an ideal testbed to study 
these processes. 
MIDI (MID-infrared Interferometer) observations recently resolved the innermost parts of 
the torus directly for the first time 
(e.~g.~\cite{Jaffe_04,Tristram_07,Tristram_09,Burtscher_09})
and found evidence for a clumpy dust distribution. 
In these proceedings, we numerically model a scenario, which is able to describe the
build-up and evolution of a nuclear gas disc or torus.  
Our simulations are put into context in Davies et al. (these proceedings).

\section{Three-dimensional hydrodynamical simulations}
\label{sec:3dmodels}

\begin{table}[b]
\begin{center}
\caption[Parameters of our standard models]{Parameters of our standard models.}
 \label{tab:param}
\begin{tabular}{lllcll}
\hline
\multicolumn{3}{c}{Hydrodynamical model}  & \hspace{1cm} & \multicolumn{2}{c}{Effective disc model} \\
\hline
Parameter & Value & Reference & & Parameter & Value\\
\hline
$M_{\mathrm{BH}}$ & $8\,\cdot 10^{6}\,M_{\odot}$ & L03 &  & $R_{\mathrm{in}}$ & $0.1\,$pc \\
$M_{*}$ & $2.2\,\cdot 10^{8}\,M_{\odot}$ & D07 & & $R_{\mathrm{out}}$ & $100.0\,$pc \\
$M_{\mathrm{gas}}^{\mathrm{ini}}$ & $1.0\,\cdot 10^{2}\,M_{\odot}$ & &  & $\delta$ & 0.2 \\
$R_{\mathrm{c}}$ & 25\,pc & G03 & & $T$ & 400\,K \\
$R_{\mathrm{T}}$ & 5\,pc &  & & $\alpha_{\nu}$ & 0.05 \\
$R_{\mathrm{in}}$ & 0.2\,pc &  & & $t_{\mathrm{cluster}}^{\mathrm{start}}$ & $50\,$Myr \\
$R_{\mathrm{out}}$ & 50\,pc &  & & $t_{\mathrm{cluster}}^{\mathrm{end}}$ & $300\,$Myr \\
$\sigma_{*}$ & 100\,km/s & D07 & & $n_{\mathrm{r}}$ & 5000 \\
$\beta$ & 0.5 &  &  & & \\
$T_{\mathrm{ini}}$ & $2.0\,\cdot 10^{6}\,$K & & & & \\
$\dot{M}_{\mathrm{n}}$ & $9.1\,\cdot 10^{-10}\,M_{\odot}/(yr\,M_{\odot})$ & J01 & & & \\
$M_{\mathrm{PN}}$ & $0.5\,M_{\odot}$ &  & & & \\
$\Gamma$ & $5/3$ & & & & \\
\hline
\end{tabular}
\end{center}
\medskip
 Mass of the black hole 
 ($M_{\mathrm{BH}}$), normalisation constant of the stellar potential ($M_{*}$), 
 initial gas mass ($M_{\mathrm{gas}}^{\mathrm{ini}}$), cluster core 
 radius ($R_{\mathrm{c}}$), torus radius ($R_{\mathrm{T}}$), inner radius ($R_{\mathrm{in}}$), 
 outer radius ($R_{\mathrm{out}}$),
 stellar velocity dispersion ($\sigma_{*}$), exponent of the angular momentum
 distribution of the stars ($\beta$),
 initial gas temperature ($T_{\mathrm{ini}}$), normalised mass injection rate ($\dot{M}_{\mathrm{n}}$), 
 mass of a single injection ($M_{\mathrm{PN}}$) 
 and adiabatic exponent ($\Gamma$), 
 thickness of the disc ($\delta$ = disc scale height / radius of the disc),
 gas temperature ($T$), alpha viscosity parameter ($\alpha_{\nu}$), 
 age of the nuclear star cluster at the beginning of the
 simulations ($t_{\mathrm{cluster}}^{\mathrm{start}}$) and at the end
 ($t_{\mathrm{cluster}}^{\mathrm{end}}$) and resolution of the simulation ($n_{\mathrm{r}}$).
 The references are: L03 \cite{Lodato_03},
 D07 \cite{Davies_07}, G03 \cite{Gallimore_03} and J01 \cite{Jungwiert_01}. 
\end{table}

\begin{figure}[t!]
\includegraphics[width=1.0\linewidth]{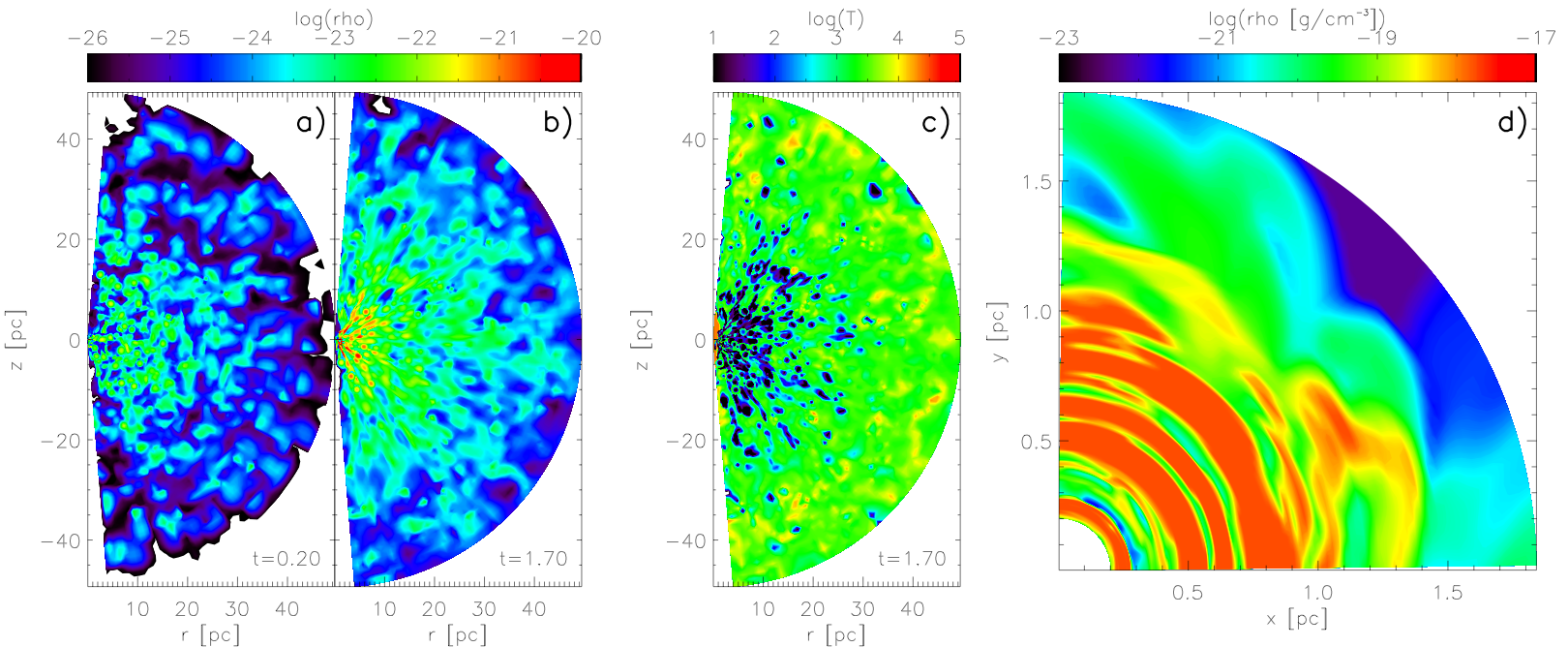}
\caption{Snapshots of the density distribution in a meridional
  plane of our 3D hydrodynamical standard model after (a) 0.2 orbits 
  (corresponding to $7 \cdot 10^4$\,yr) and (b) 1.7 orbits
  (approximately $6 \cdot 10^5$\,yr). Panel (c) shows the
  temperature distribution in the same meridional plane after 1.7 orbits. 
  Panel (d) is a zoom into the innermost part, showing the nuclear disc
  component.}
\label{fig:den_temp_inndisk} 
\end{figure}

Recently, high resolution observations with the help of the near-infrared
adaptive optics integral field spectrograph SINFONI at the VLT proved the
existence of massive and young nuclear star clusters in the centres of a
sample of Seyfert galaxies (\cite{Davies_07} and Davies et al., these proceedings). 
With the help of three-dimensional high resolution hydrodynamical
simulations with the {\sc Pluto} code, we follow the evolution of
such clusters. The gas ejection of their stars provide both, 
material for the obscuration within the mentioned {\it Unified Scheme
of AGN} and a reservoir to fuel the central, active region and it additionally 
drives turbulence in the interstellar medium on tens of parsec scales.
We start our simulations after the very violent phase of supernova type\,II 
explosions and fast winds. 
In order to enable a direct data comparison, we constrain our 
input parameters with observed values of the well-studied nearby Seyfert galaxy
NGC\,1068. A summary of the parameters used for the simulations shown in 
these proceedings is given in Table\,\ref{tab:param}.
After approximately 50\,Myrs following the starburst, the mass loss of the 
newly born stellar population is dominated by slow winds and the ejection of planetary nebulae 
respectively. Then, the input of energy into the ambient medium is low enough to enable 
transport of gas towards the centre allowing for the triggering of activity there
(see also Davies et al., these proceedings).
A similar time delay between the nuclear starburst and the onset of nuclear activity has
also been found observationally, leading to the same conclusions as our work (\cite{Davies_07}).
We model the mass-loss of the stellar population as the ejection of single expanding clumps 
of gas, which we give the velocity of the underlying stellar distribution at the point of 
emission, made up of a rotation component and a random component\footnote{These nuclear star clusters
possess a significant velocity dispersion and rotate with sub-Keplerian velocity.}.
These blobs of gas -- confined by cooling instability -- merge to form larger entities, thereby 
dissipating a fraction of their initial turbulent motions and get transported towards the centre, 
forming slightly elongated filaments. 
These processes lead to a vertically wide distributed clumpy or filamentary inflow of gas on
tens of parsec scale (see Fig.~\ref{fig:den_temp_inndisk}a,b,c), 
whereas a turbulent and very dense disc builds up on
the parsec scale (see Fig.~\ref{fig:den_temp_inndisk}d). 
A similar two-component structure has been found in interferometric observations with the 
help of the MIDI instrument as well in NGC\,1068 \cite{Raban_09} and the Circinus galaxy \cite{Tristram_07}. 

Being computationally very expensive simulations and as we currently are unable to 
take all physical processes happening in the innermost few parsec 
into account in the three-dimensional hydrodynamical calculations, we treat the 
inner disc separately with the help of an effective disc model, as described in 
Sect.\,\ref{sec:eff_discmodel}.

\section{Effective disc modelling}
\label{sec:eff_discmodel}

In order to capture the relevant physics in the inner region, we treat 
this disc separately in a one-dimensional, axisymmetric simulation, which also enables a
direct comparison with observations. Due to the generally unknown
physical mechanism to generate angularm momentum and mass transfer in the disc,
we concentrate on the effects of a parametrised turbulent viscosity 
and additionally take star formation into account.
Following \cite{Pringle_81} and \cite{Lin_87}, the viscous evolution of such
a disc can be described by the following differential equation:

\begin{eqnarray}
\label{equ:dgl}
\frac{\partial}{\partial t} \Sigma(t,R) + \frac{1}{R} \, \frac{\partial}{\partial
  R} \, \left[\frac{\frac{\partial}{\partial R} \left(
      \nu_{\alpha} \, \Sigma(t,R) \, R^3 \, \frac{d \Omega(R)}{d R}
    \right)}{\frac{d}{d R} \left( R^2 \, \Omega \right)} \right]
= S, 
\end{eqnarray}

where $\Sigma$ is the gas surface density within the disc, $R$ is the radial distance from 
the centre, $t$ the time, $\nu_{\alpha}$ the assumed alpha viscosity, $\Omega$ the Keplerian 
rotation frequency and $S$ includes a source term for the mass input 
from our three-dimensional hydrodynamical simulations (Sect.\,\ref{sec:3dmodels})
and a sink term, taking star formation into account
with the help of the Kennicutt-Schmitt law.
The infalling material is placed according to its angular momentum in a Keplerian 
rotating disc structure (Fig.~\ref{fig:input_mass_scal}a). Gas consumed into stars 
is removed from the simulations and will no longer 
participate in the dynamical evolution.
We solve Equ.~\ref{equ:dgl} numerically with the help of  
{\sc Matlab}'s\footnote{http://www.mathworks.com/products/matlab/} pdepe solver.
This two-stage modelling process enables us to directly compare our results to
observed properties.

\section{Results and data comparison}

\begin{figure}[b!]
\centering
\includegraphics[width=0.55\linewidth]{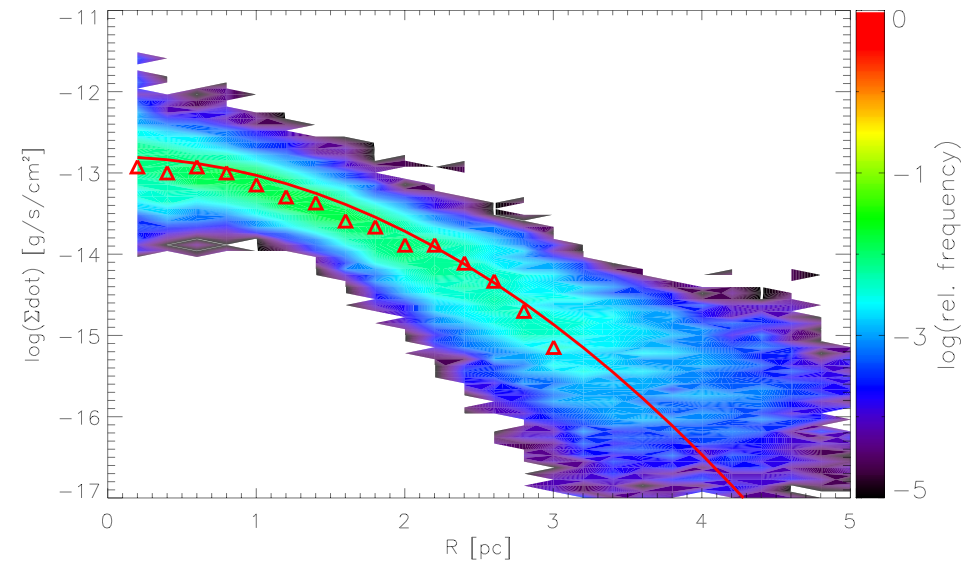}
\hfill
\includegraphics[width=0.4\linewidth]{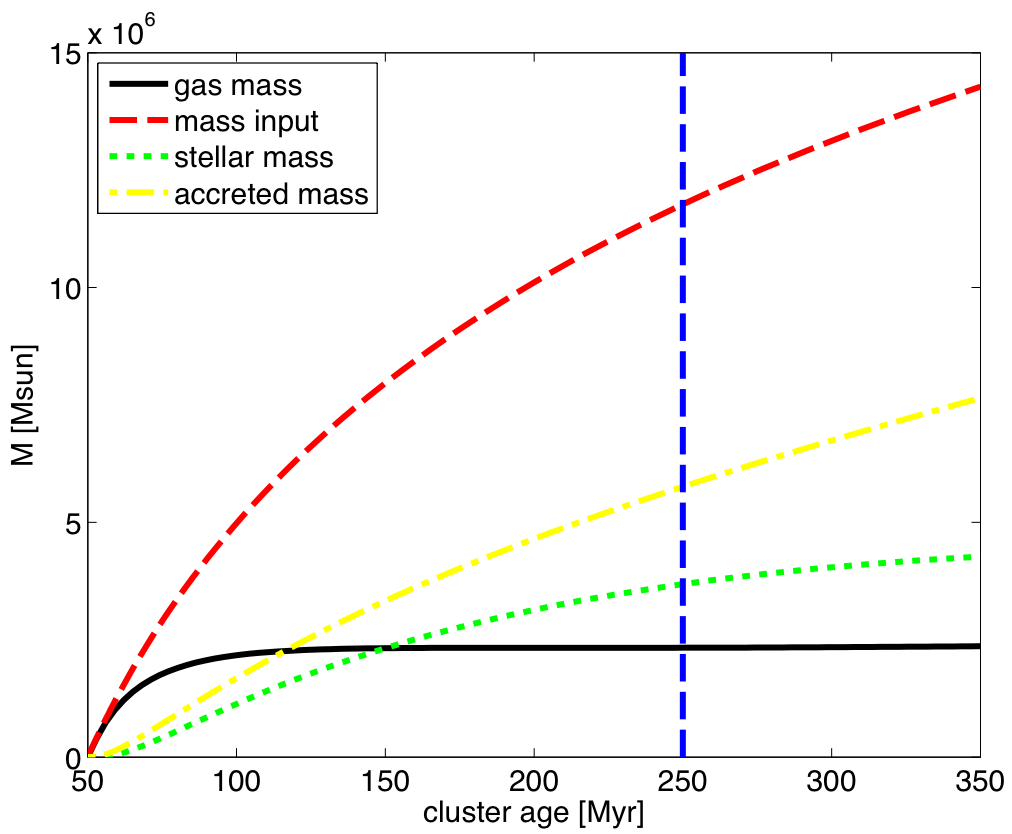}
\caption{{\bf (a)} Mass input into our effective disc model,
  shown as the growth of the gas surface density 
  at different radii, 
  flowing in through a sphere with radius of $2.5\,$pc from our 3D {\sc Pluto} turbulent torus simulation. 
  The input radii are set according to an angular momentum
  distribution of a Keplerian disc. Shown is the logarithm of a histogram for 100
  timesteps with a resolution of approximately 3400\,yr between an evolution
  time of 0.7\,Myr and 1.0\,Myr, where the 3D hydro simulation is already in an
  equilibrium state. The triangles denote the maxima of each radial bin, the
  parabola is the parametrisation used as source term in the effective disc simulations. 
  {\bf (b)} Comparison of mass contributions of the various components of our
  1D effective disc model for an alpha viscosity parameter of 0.05 and a gas temperature of 400\,K.
  The vertical dashed line denotes the estimated current age of the nuclear star
  cluster in NGC\,1068.
% figure produced with:
% ex-srv4:~/data/routines/IDL/PLUTO_IDL/disk_matlab_input_new_cool.pro
% in ex-srv4:~/data/LRZ_copy/2009/tor_highres_log_copy_cool/
}
\label{fig:input_mass_scal}       
\end{figure}

\begin{figure}[b!]
\begin{center}
\includegraphics[width=0.45\linewidth]{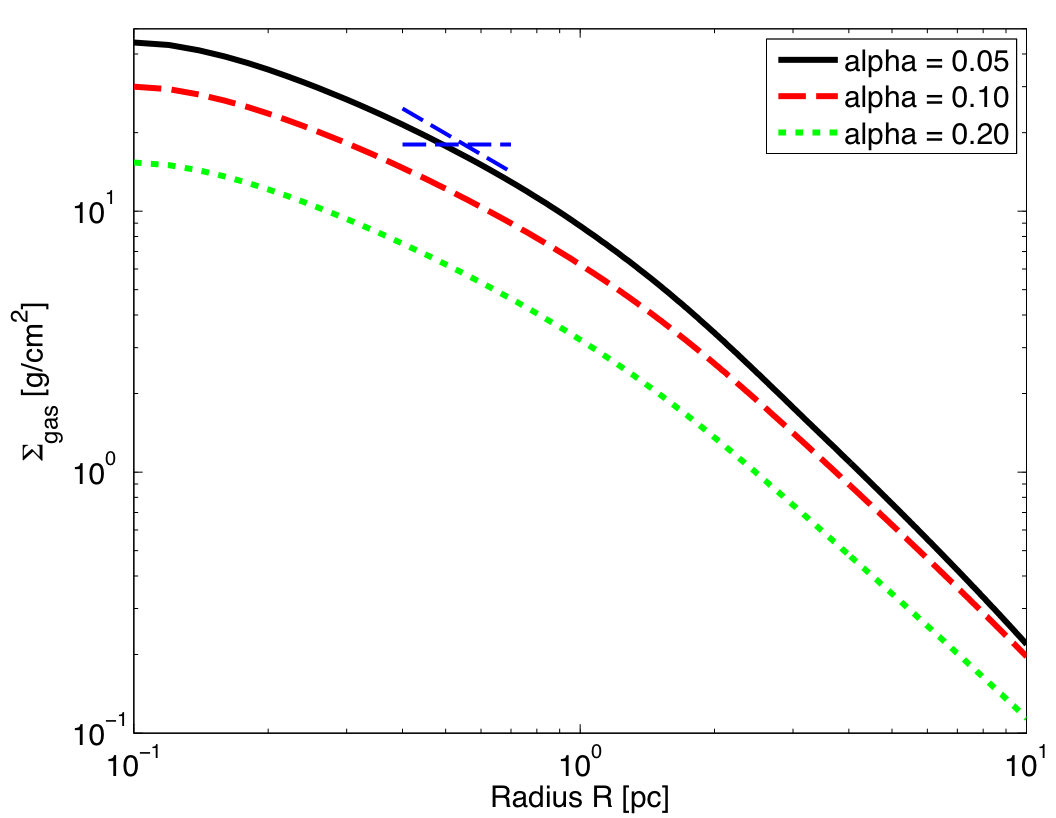}
\hfill
\includegraphics[width=0.45\linewidth]{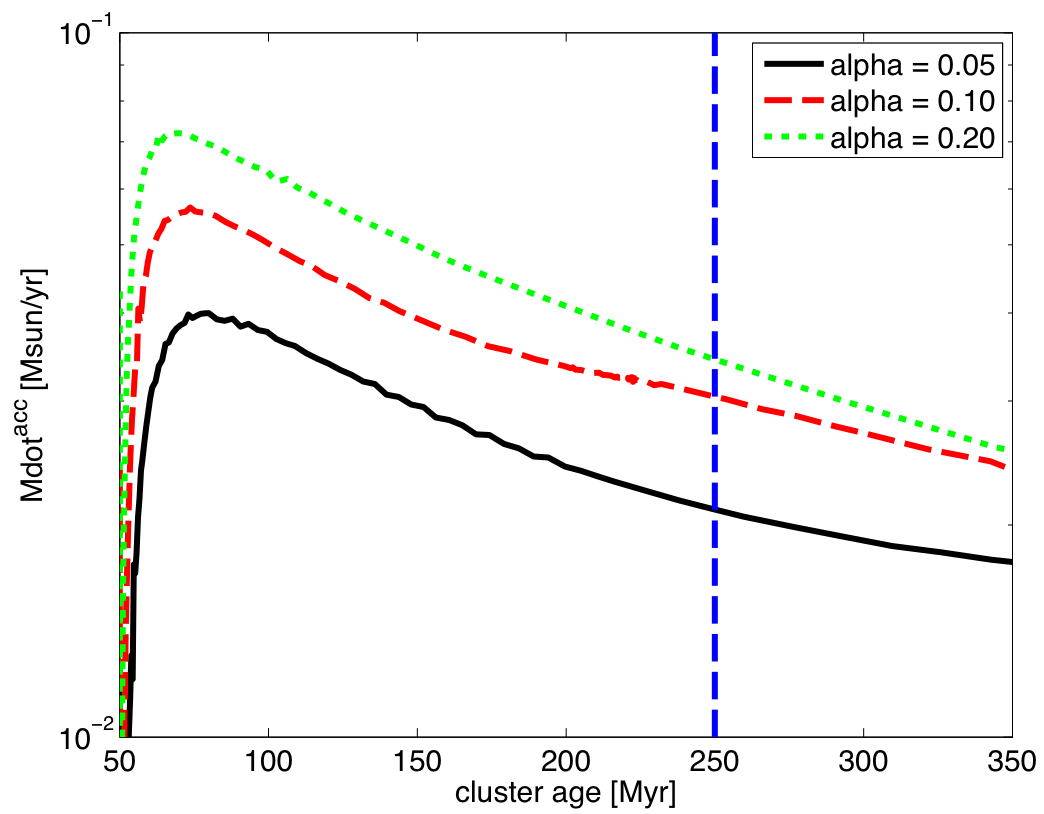}
\caption{{\bf (a)} Final gas density distribution at a cluster age of
  250\,Myr. Superposed as dashed lines are the two limiting curves for
  the range of possible surface density distributions as determined from MIDI
  observations \cite{Kishimoto_09}.
  {\bf (b)} Total accretion rate of gas through the inner boundary of the domain.  
}
% figures produced with:
% plot_study.m
% in mschartm@ex-sol:~/calculations/disk_evolution/paper_const_scaleheight_revised/alpha_study_kepler/
\label{fig:gasfinal_mdot}
\end{center}
\end{figure}

Fig.~\ref{fig:input_mass_scal}b shows the comparison of the mass contributions
of the various components from the 1D modelling for an assumed alpha viscosity
parameter of $\alpha = 0.05$ and a gas temperature of 400\,K. 
The input parameters of the simulations have been constrained by observations
of the nearby Seyfert~2 galaxy NGC~1068 (see Table\,\ref{tab:param}).
At the current age of the starburst in NGC~1068 of approximately 250\,Myr, 
the largest fraction of the integrated gas mass introduced into 
the disc simulations (given by the dashed line
in Fig.~\ref{fig:input_mass_scal}b) has been accreted with the help 
of viscous processes (dash-dotted line). A slightly smaller amount has 
been transformed into stars and a few times $10^6\,M_{\odot}$ of gas remains in
the disc (solid line), which has been built up within a few million years.  
The latter is in good comparison with the disc mass \cite{Kumar_99} finds by 
fitting a clumpy disc model to the maser detections.
Fig.\,\ref{fig:gasfinal_mdot}a shows the surface density distribution for three simulations
with varying strength of the viscosity: $\alpha=0.05$, 0.1 and 0.2. For comparison,
the range of slopes of the density distribution as inferred from MIDI observations
of a sample of nearby Seyfert galaxies is overplotted as dashed line segments. 
The sizes of the corresponding dust distributions we derive from these curves amount
to 0.8 to 0.9\,pc, which is remarkably similar to values obtained with the
help of the MIDI instrument of approximately 0.7\,pc and to the estimated 
extent of the maser disc (0.65 to 1.1\,pc).
Finally, the total accretion rate through the inner boundary of our disc 
simulations is plotted in Fig.~\ref{fig:gasfinal_mdot}b for the $\alpha$ parameter study 
discussed above. Larger values of $\alpha$ clearly lead to a larger mass accretion
rate and at the current age of the nuclear stellar cluster, the derived values
agree well with typical accretion rates of Seyfert galaxies of a few times 
$10^{-3}$ to $10^{-2}\,M_{\odot}/$yr (\cite{Jogee_06}).
However, NGC\,1068 seems to be in a heavily accreting state, which can be accomodated in our
model only, when assuming e.\,g.\,clumpy accretion.

\section{Concluding remarks}

In these proceedings, we show that evolving stars from a massive
and young nuclear star cluster, as found in nearby Seyfert galaxies provide
enough gas to assemble a parsec-sized nuclear gas disc. 
To this end, we combine a three-dimensional treatment of the mass-loss from an 
evolving nuclear star cluster with 
a simplified model for the 
innermost parsec scale region, where a nuclear disc builds up.
As far as possible, we derive input parameters of our simulations from 
observations of the nearby and well-studied Seyfert\,2 galaxy 
NGC\,1068.
This two-stage analysis enables us to (i) do a 
long term evolution study, (ii) link the tens of parsec scale region 
of galactic nuclei (observed with the SINFONI instrument) to the sub-parsec
scales (probed by MIDI and in water maser emission) and
(iii) test our model directly with a large number of observational 
results. 
At the current age of its nuclear starburst of 250\,Myr, our simulations yield 
disc sizes of the order of 0.8 to 0.9\,pc, gas masses of $10^6\,M_{\odot}$ and mass
transfer rates of $0.025\,M_{\odot}/\mathrm{yr}$ through the inner rim of the
disc in good comparison with observed disc and torus properties. 
On basis of these comparisons, we conclude that the proposed scenario seems to
be a reasonable model and shows that 
nuclear star formation activity and subsequent AGN activity are intimately
related.

\vspace{1cm}

{\bf Acknowledgment:}\\
\newline
Part of the numerical simulations have been carried out on the SGI Altix 4700 HLRB\,II of 
the Leibniz Computing Centre in Munich (Germany).

\end{document}